\documentclass[aps,prb,twocolumn,showpacs]{revtex4}
\usepackage[dvips]{graphicx}
\usepackage{wrapfig}
\usepackage{psfrag}

\begin{document}

\title{2D SNS junction with Rashba spin-orbit interaction}

\author{Ol'ga Dimitrova and M. V. Feigel'man}

\affiliation{L.D.Landau Institute for Theoretical Physics, Kosygin
str.2, Moscow, 119334, Russia}

\date{\today}

\begin{abstract}
The effect of Rashba spin-orbital interaction upon supercurrent in
S-2DEG-S  proximity junctions  is investigated in the clean limit.
Generalization of Beenakker's formula for Andreev levels to the
case of spin-orbital scattering is presented. Spin-orbit - induced
splitting of Andreev bound-states is predicted for a junction of
infinite width, and with non-vanishing normal backscattering at
S/N interfaces. Semiclassical average of the Josephson current is
however insensitive to the Rashba coupling as long as
electron-electron interaction in 2DEG is neglected.

\end{abstract}

\pacs{74.50.+r, 74.60.Jg, 71.70.Ej}

\maketitle

\section{Introduction}

Josephson junctions of two superconductors via two-dimensional
electron gas, (most usually implemented in the Nb/InAs/Nb
structures) were actively studied both experimentally and
theoretically, cf. e.g.~\cite{1,2,3,4,5,6}. Generic feature of all
these devices is strong reduction of the experimentally measured
product $I_cR_N$ with respect to the theoretical predictions. In
particular, this discrepancy is known for short junctions with
high-quality S/N interfaces, demonstrated by measurement of a
non-sinusoidal current-phase relation~\cite{6}. At the
temperatures much below $T_c$, the parameter $I_cR_N \approx 0.22
mV$ was measured in Ref.~\onlinecite{6}, to be compared with the
Niobium superconductive gap $\Delta \approx 1.5 meV$. Thus it
seems natural to look for some  effects which were not taken into
account in the existing theory, cf.
e.g.~\cite{golubov,chrestin94}, but could be responsible for such
a drastic suppression of the critical current.

An obvious candidate to be explored is the Rashba spin-orbital
interaction~\cite{Rashba} $H_{R} = \alpha[{\bf \sigma}\times{\bf
p}] \cdot {\bf n}$  known to exist in the 2DEG structures due to
up-down asymmetry of the quantum well (here $\bf n$ is the unit
vector normal to the plane of the 2DEG). In the InAs
heterostructures this term is especially large (cf.
Ref.~\onlinecite{Nitta}), leading to the band splitting $\Delta_R
= 2\alpha p_F \approx 5 meV$, i.e. considerably larger than the
Niobium superconductive gap. Therefore  it seems natural that the
account of the Rashba term might be important for analysis of the
Josephson current in these devices.  We also note in this relation
the paper~\cite{Universal}, where it was shown that persistent
currents in mesoscopic metal rings should be modified strongly by
SO coupling - which seems to point out the existence of similar
effect upon the Josephson current.

However, it is frequently assumed that the spin-orbital
interaction cannot influence proximity effect  in superconductive
structures, since it respects time-reversal invariance. This
argument is not valid, however, when the critical Josephson
current is considered,
 since the presence of a current already breaks down the time-reversal
symmetry. More detailed arguments seem to come from recent
papers~\cite{bezuglyi,krive}, where the influence of both Rashba
coupling and Zeeman magnetic field upon the critical current of
S-N-S junctions was considered.  In both these papers it was found
that in the absence of the Zeeman term the Rashba interaction (if
treated within the simplest model of equal Fermi velocities on
both chiral branches) totally cancels out from the equations for
Andreev levels. We will show, however, that this cancellation is
not generic; rather it is due to different simplifications used in
the papers mentioned: a model of completely transparent S/N
interfaces was employed in Ref.~\onlinecite{bezuglyi}, and a
purely one-dimensional model used in Ref.~\onlinecite{krive}.

It will be shown below that in the general case, when some normal
backscattering at an arbitrary angle of incidence occurs at the
S-N boundaries, the spin-orbital coupling does affect the energies
of the Andreev levels and the supercurrent they carry on. We will
show that the effect of SO interaction can be understood as being
due to modification  of the transmission channels defined by the
scattering matrix $\check{S}$, which describes the junction
properties in the normal state. For a model junction with infinite
length (or periodic boundary conditions) in the direction
transverse to the supercurrent, spin-orbit splitting of
transmission eigenvalues is found, which results in splitting of
each  Andreev level into pair  of spin-polarized levels, with
phase-dependent energy difference $\delta E(\chi)$. Note, that
$\delta E(0)=0$,  in agreement with the time-reversal invariance
restored in the absence of the phase bias. The idea that the
Andreev levels can be spin-splitted due to the SO coupling was
proposed in Ref.~\onlinecite{SN} for a narrow (few-channel)
junction. The SO effect we discuss in the present paper is
different from the one presented in Ref.~\onlinecite{SN}.

In this paper we consider the simplest two-dimensional model of
ballistic S-2DEG-S junction (cf.
e.g.~Ref.~\onlinecite{chrestin94}) of infinite width in the
lateral direction transverse to the current flow, see. Fig.~1.  We
neglect possible potential barriers at the S/N interfaces,
assuming that the normal backscattering is due to Fermi-velocity
mismatch only, and consider ballistic electron propagation along
2D structure between superconductive terminals. In the Sec. II of
the paper we show that in the short junction limit (junction
length $L \ll \xi_0 = \hbar v_F/\Delta$, where $v_F$ is the
Fermi-velocity of 2DEG) the positions of the Andreev levels can be
expressed via the transmission eigenvalues $\cal T$ of the full
scattering matrix $\check{S}$ in precisely the same way as was
found by Beenakker ~\cite{Bee91} for junctions with
spin-independent scattering. Then in Sec. III we present
calculations of the scattering matrix $\check{S}$ for the simplest
two-dimensional model of ballistic S-2DEG-S junction (cf. e. g.
Ref.~\onlinecite{chrestin94}) of infinite width in the lateral
direction transverse to the current flow, see. Fig.~1. We
demonstrate explicitly spin-splitting of the transmission
probabilities ${\cal T}_\pm(p_y)$ for the transmission channels
characterized by the momentum component $p_y$. We show then, that
the distribution function for the transmission probabilities
${\cal P}({\cal T})$ coincides with the one discussed by Melsen
and Beenakker~\cite{MelBee} in the absence of spin-orbital
coupling. In the Sec. IV  we derive expression for the Josephson
current of a short junction and demonstrate that {\it average}
current is insensitive to the Rashba coupling.  In Sec.V we go
beyond the short-junction limit: we derive an equation for
spin-splitted Andreev levels for the junction with arbitrary
$L/\xi_0$ ratio and demonstrate that their contribution to the
average (semiclassical) supercurrent is insensitive to the SO
coupling. Section VI is devoted to the discussion of applicability
of our results to S-2DEG-S junctions of finite width and  possible
ways to detect spin-splitted Andreev levels. Finally, in Sec. VII
we present our conclusions and discuss open problems; in
particular, it is proposed that the account of electron-electron
interaction {\it together} with Rashba coupling might be able to
explain Josephson current.

\psfrag{-L/2}{\kern-5pt\lower0pt\hbox{{$-L/2$}}}
\psfrag{L/2}{\kern-2pt\lower0pt\hbox{{$L/2$}}}
\psfrag{N}{\kern0pt\lower0pt\hbox{{${{\bf {n}}}$}}}
\psfrag{A1}{\kern0pt\lower0pt\hbox{{$m, v_s$}}}
\psfrag{A2}{\kern-9pt\lower0pt\hbox{{$m_n, v_n$}}}
\psfrag{A3}{\kern-12pt\lower0pt\hbox{{$m, v_s$}}}
\psfrag{0}{\kern0pt\lower-2pt\hbox{{$0$}}}
\psfrag{Y}{\kern-5pt\lower0pt\hbox{{ $y$}}}
\psfrag{X}{\kern-1pt\lower-2pt\hbox{{ $x$}}}
\psfrag{A}{\kern-2pt\lower-3pt\hbox{{ $\varphi$}}}
 \begin{figure}[tbp]
 \hspace{15mm}
 \includegraphics[width=0.45\textwidth]{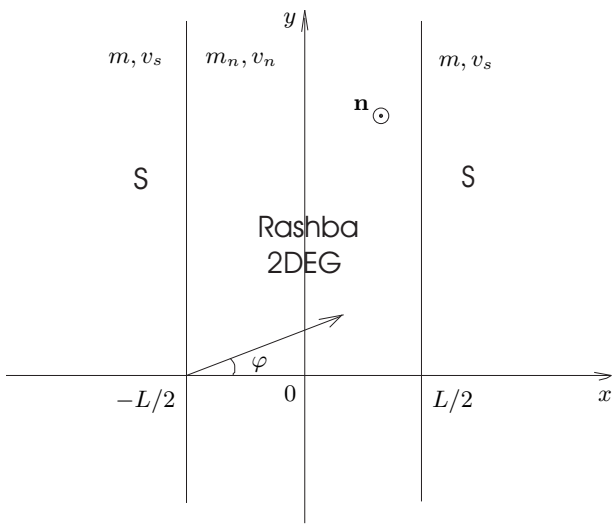}
 \caption{{Two-dimensional model of a
superconductor/Rashba 2DEG/superconductor Josephson junction
infinite in direction perpendicular to the current (along $y$
axis). The Rashba 2DEG region has thickness $L$; $m$/$m_n$ is the
effective mass and $v_s$/$v_n$ is the Fermi velocity in the
S/2DEG; $\varphi$ is the angle between the velocity direction of a
quasiparticle and the $x$ axis in the 2DEG region; ${{\bf n}}$ is
a unit vector normal to the plane of the 2DEG. }} \label{}
 \end{figure}

\section{Spectrum of Andreev levels}

The excitation spectrum consists of the positive eigenvalues of
the Bogolyubov-de Gennes (BdG) equation:
\begin{eqnarray}
\epsilon_{\nu}u^{\alpha}={\left[\xi+U\right]^\alpha}_\beta
u^\beta+\Delta
v^{\alpha}\nonumber \\
\epsilon_{\nu}v^{\alpha}=-{\left[\xi^*+U^*\right]^\alpha}_\beta
v^\beta+\Delta^*u^\alpha,
\end{eqnarray}
where
${\left(U^*\right)^\alpha}_\beta=\hat{g}^{\nu\alpha}
(U(\nu,\mu))^*\hat{g}_{\mu\beta}$,
$\alpha,\beta$ are spinor indices, $\hat{g}=i\hat{\sigma}_y$ is a
metric tensor in spin space; $\xi=\frac{{\bf p}^2}{2 m}-E_F$ is
the kinetic energy of a quasiparticle (energies are measured
relative to the Fermi energy);
\begin{eqnarray}
u(\vec{r})^{\alpha}=\left(%
\begin{array}{c}
  u(\vec{r}\uparrow) \\
  u(\vec{r}\downarrow) \\
\end{array}%
\right),\quad u_{\alpha}=\hat{g}_{\alpha\beta}u^{\beta},\quad
u^{\alpha}=u_{\beta}\hat{g}^{\beta\alpha},
\nonumber \\
U(\sigma,\mu)={U^{\sigma}}_\mu.
\end{eqnarray}
In our model in the normal region (mapped on Fig. 1 as ''Rashba
2DEG``) the operator $U=\alpha[{\bf \sigma}\times{\bf p}] \cdot
{\bf n}$ is the spin-orbit interaction, which preserves the
time-reversal invariance. In the superconductors the Rashba term
is absent, $U=0$. The superconducting gap $\Delta$ is assumed to
be a step-like function: equal to zero in the normal region, and
its' modulus $|\Delta|$ constant and equal in both
superconductors.

Equation which relates the excitation spectrum of the Josephson
junction to the scattering matrix in the normal state $\check{S}$,
was derived in Ref.~\onlinecite{Bee91}:
\begin{eqnarray}\label{Andreevdet1}
\det[1-r_{he}\check{S}_e(\epsilon)r_{eh}\check{S}_h(\epsilon)]=0,
\end{eqnarray}
where
\begin{eqnarray}
&&r_{he}=\gamma r_A,\quad r_{eh}=\gamma r_A^*,\quad r_A=\left(%
\begin{array}{cc}
  e^{i\chi/2} & 0 \\
  0 & e^{-i\chi/2} \\
\end{array}%
\right),\nonumber \\
&&\gamma=e^{-i\arccos(\epsilon/\Delta)},
\end{eqnarray}
$r_{he}$ is the Andreev reflection matrix for $e\rightarrow h$
scattering in the space of channels incident (reflected) on the
left and right NS boundary, $\pm\chi/2$ are the phases of the left
(right) superconductor, $\check{S}_{e(h)}$ is the electron (hole)
scattering matrix of the normal state.

When no spin-dependent scattering is present, normal scattering
matrix $\check{S}_e$ is trivial in spin space, i. e. proportional
to the unit matrix $\hat{\sigma}_0$. Further on, within the short
junction limit $L \ll \xi_0$, scattering matrices
$\check{S}_{e,h}$ do not depend on energy, moreover $\check{S}_h =
\check{S}^*_e$. Therefore Eq.(\ref{Andreevdet1}) can be
transformed to an explicit solution~\cite{Bee91} for
spin-degenerate Andreev levels, $\epsilon_j = \pm \sqrt{1 - {\cal
T}_j\sin^2\frac{\chi}{2}} $, where ${\cal T}_j$ is the $j$-th
eigenvalue of the transmission probability matrix $\hat{T}^\dagger
\hat{T}$ (eigenvectors of this matrix define scattering channels).
Below we show that solution of the same kind can be obtained when
the spin-orbital scattering is present.

In the presence of spin-orbital interaction the scattering matrix
in~(\ref{Andreevdet1}) becomes spin-dependent but still obeys
time-reversal invariance. It makes possible to generalize
Beenakker's derivation for Andreev levels in short
junction~\cite{nms} using the following set of relations for the
$\check{S}$-matrix:
\begin{eqnarray}
\check{S}\check{S^\dagger} = 1,     \qquad   \qquad
\check{S}^T(-p_y)=\hat{g}^T\check{S}(p_y)\hat{g}, \,\,\,  \nonumber \\
\check{S}_h(\epsilon,p_y)=\hat{g}^T\check{S}^*_e(-\epsilon,-p_y)
\hat{g},
\label{3cond}
\end{eqnarray}
where
the superscript $^T$ means full matrix transposition. The first
relation in (\ref{3cond}) is just a unitarity condition, the
second one follows from the time-reversal invariance (we used here
a transformation of the wavefunctions upon time-reversal,
$\psi^{t-r}(p_y) = \hat{g}\psi^*(-p_y)$). Finally, the third
relation in (\ref{3cond}) is due to a special symmetry of the
Bogolyubov-de Gennes equations:  $\psi_h(\epsilon,p_y) =
\hat{g}^T\psi^*_e(-\epsilon,-p_y)$. It is important to note the
sign change of the parameter $p_y$ in the second and the third
relations above: when all scattering states are characterized by a
conserved momentum ($p_y$), the time-conjugation  operation
contains complex conjugation and  $p_y \to -p_y$ inversion, since
time reversal of the scattering matrix should change the sign of
$p_x$ only, while keeping $p_y$ intact. In other terms, the
additional $p_y \to -p_y$ operation is needed due to the use of
scattering channels characterized by {\it complex} eigenfunctions
$\propto e^{ip_y y}$. Usually in calculations of this kind a {\it
real} basis of transmission channels is used, in which case such
an additional operation would be absent.

Making use of relations (\ref{3cond}), one can transform
Eq.~(\ref{Andreevdet1}) into the form:
\begin{equation}\label{Andreevdet2}
\det\left[\frac{1}{\gamma}\hat{g}^T\check{S}_e^*(\epsilon,p_y)
\hat{g}r_A^*-\gamma
r_A^*\hat{g}^T\check{S}_e^*(-\epsilon,p_y)\hat{g}\right]=0.
\end{equation}
For a short contact $L\ll \xi_0$, we neglect the energy-dependence
 of the scattering matrix in Eq.~(\ref{Andreevdet2}),
and obtain a second order equation with respect to $\epsilon^2$,
which results in the following solution:
\begin{equation}\label{beenakker}
\epsilon_{s,\eta}(p_y)= \eta\Delta\sqrt{1-{\cal
T}_s(p_y)\sin^2{\frac{\chi}{2}}},
\end{equation}
where $\eta = \pm$ and ${\cal T}_s(p_y)$ are transmission
probabilities - eigenvalues of the matrix
$\hat{T}^{\dagger}\hat{T}$, depending on the spin index $s=\pm$
and the conserved momentum $p_y$. In general, ${\cal T}_+(p_y)
\neq {\cal T}_-(p_y) $, thus four non-degenerate Andreev levels
correspond to each $p_y$ value, as shown in Fig.3 below.
Note however, that full family of Andreev levels still contains
pair-wise degeneracy within our model. Namely, degeneracy exists
between states with $p_y = \pm |p_y|$.
Below we consider a specific
example of scattering problem relevant to S-2DEG-S structures, and
calculate ${\cal T}_s(p_y)$ eigenvalues.

\section{S-matrix and transmission eigenvalues}

We are interested in the study of specific spin-orbital effects
and thus will consider the simplest  model of S/N boundaries,
assuming that normal electron reflection is due to Fermi velocity
mismatch only,  $v_s \neq v_n$  (here $v_s$ and $v_n$ are the
Fermi velocities in the superconductive metal and in the 2DEG
correspondingly). Additional source of reflection due to an
effective potential barrier at the interface (cf. e. g.
Ref.~\onlinecite{chrestin94}) can be present, but does not affect
our results qualitatively. Since the effective mass $m_n$ in the
2DEG differs strongly from the effective mass $m$ in the metallic
superconductor (typically, $m_n/m \approx 0.03$ for 2D structures
with InAs), the difference of these masses should be taken into
account explicitly. Our first goal now is to find
reflection/transmission amplitudes on single S/N interfaces (for
the normal state of the superconductive metal S). We will follow
Ref.~\onlinecite{Finkel'stein}, and use continuity equations which
follow from the Schrodinger equation with space-dependent mass
$m(x)$ and spin-orbital parameter $\alpha(x)$:
\begin{equation}
\left[\frac{\hat{p}_x}{m(x)} - \alpha(x)\right]\Psi\mid^S_N = 0 \,
, \qquad \Psi\mid^S_N = 0, \label{contin}
\end{equation}
where $F\mid^S_N$ denote  $F(x= -\frac{L}{2} + 0) - F(x=
-\frac{L}{2} - 0)$ for the left interface (cf. Fig. 1) and
similarly for the right interface located at $x=L/2$. Further on,
$P_F = mv_s$ and $p_F = m_nv_n $ are the Fermi-momenta in the S
metal and in the 2DEG correspondingly; usually $p_F \ll P_F$,
whereas $v_s$ and $v_n$ are of the same order of magnitude. Below
we will assume that the parameter $\alpha/v_n \ll 1$ measuring the
relative strength of the Rashba interaction is small in comparison
with the Fermi-velocity mismatch, i.e. $\alpha \ll |v_s-v_n|$.
Under this condition, reflection amplitudes at each of the S/N
boundaries are determined by the ratio $v_n/v_s$ only. Then the
amplitudes of reflection and transmission are trivial in the spin
space, e.g. $\overrightarrow{r}^{\alpha\beta}_1 =
\delta^{\alpha\beta}\overrightarrow{r}_1 $, and so on. For an
incident wave incoming from $x=-\infty$, the amplitudes of
reflection and transmission on the left (1) interface, are:
\begin{equation}\label{TRsingle}
\overrightarrow{r}_1 = \frac{w-1}{1+w} \, , \qquad
\overrightarrow{t}_1 = \frac{2}{1+w}, \label{rt1}
\end{equation}
where $w = v_{nx}/v_{sx}$ is the ratio of the $x$-components of
the electron velocities, with $v_{nx} = v_n\cos\varphi$ and
$v_{sx} = \left[v_s^2 -
(\frac{m_n}{m})^2v_n^2\sin^2\varphi\right]^{1/2} \approx v_s$.
Here $\varphi$ is the angle between the velocity direction and the
$x$ axis in the 2DEG; note that $v_{sx}$ is very close to $v_s$
for any angle $\varphi$, since $(m_n/m)^2 \ll 1$. Other
reflection/transmission amplitudes are determined as follows:
\begin{eqnarray}
\overleftarrow{r}_2 = \overrightarrow{r}_1 \, , \qquad
\overleftarrow{t}_2 = \overrightarrow{t}_1,
\nonumber \\
\overleftarrow{r}_1 = \overrightarrow{r}_2 = -
\overrightarrow{r}_1  \, , \qquad \overleftarrow{t}_1 =
\overrightarrow{t}_2 = \frac{2w}{1+w}.
\end{eqnarray}

The total scattering matrix $\check{S}$ of the S/Rashba 2DEG/S
junction in the normal state, formed out of the ``single
boundary'' amplitudes, Eq.~(\ref{TRsingle}), reads (similar
equations can be written for $\hat{T}_2$ and $\hat{R}_2$):
\begin{eqnarray}\label{S_matrix}
&&\hat{T}_1=\overrightarrow{t}_2\hat{S}^r\left[1-
\overleftarrow{r}_1(\hat{S}^l)^{-1}\overrightarrow{r}_2
\hat{S}^r\right]^{-1}
\overrightarrow{t}_1,\nonumber \\
&&\hat{R}_1=\overleftarrow{t}_1(\hat{S}^l)^{-1}
\overrightarrow{r}_2\hat{S}^r
\left[1-
\overleftarrow{r}_1(\hat{S}^l)^{-1}
\overrightarrow{r}_2\hat{S}^r\right]^{-1}
\overrightarrow{t}_1+\nonumber
\\
&&+\overrightarrow{r}_1,
\nonumber \\
\end{eqnarray}
where $\hat{R}$ and $\hat{T}$ are the reflection and the
transmission blocks of the scattering matrix:
\begin{eqnarray}
\check{S}=\left(%
\begin{array}{cc}
  \hat{R}_1 & \hat{T}_2 \\
  \hat{T}_1 & \hat{R}_2 \\
\end{array}%
\right);
\end{eqnarray}
the index $1$ in the amplitudes $\hat{R}$, $\hat{T}$ in
Eq.~(\ref{S_matrix}) means that the equations are written for the
case of an electron propagating from left to right. Matrices in
the spin space $\hat{S}^{r(l)}$ describe spin rotation during the
electron propagation across 2DEG region with the Rashba coupling
between the two S/N boundaries. Explicit form of these matrices
can be obtained by transformation of the plane-wave eigenmodes
with definite chirality to the spin basis with definite $S_y$
projection:
\begin{eqnarray}\label{SSrl}
\hat{S}^{r} = e^{i\xi}\left[ \cos{A} -i\sin{A}\sin\varphi
\hat{\sigma}_x +
i\sin{A}\cos\varphi \hat{\sigma}_z \right], \\
\nonumber (\hat{S}^{l})^{-1} = e^{i\xi}\left[ \cos{A}
-i\sin{A}\sin\varphi \hat{\sigma}_x  - i\sin{A}\cos\varphi
\hat{\sigma}_z \right].
\end{eqnarray}
Here $\xi(\epsilon)= k(\epsilon)L$, with
$k(\epsilon)=k+m\epsilon/k$ and $k=p_F\cos{\varphi}$, is the main
semiclassical phase and $A=m_n\alpha L/\cos\varphi$ is the
additional phase due to spin rotation by Rashba coupling. Within
our approximation $\alpha/v_n \ll 1$,  the whole effect of the
Rashba coupling is contained in the phase $A$ which is not small
if length $L$ of the junction is comparable or larger than the
spin-rotation length $L_0 = \hbar/m_n\alpha$.

We define, for further convenience, a new parameter $x =
\log\frac{1+w}{1-w}$, where $w$ is defined after Eq.~(\ref{rt1}).
Then, combining Eqs.~(\ref{S_matrix}) and (\ref{SSrl}), we obtain
a transmission matrix in the form:
\begin{eqnarray}\label{T}
&&\hat{T}_1=T_0+T_1\hat{\sigma}_x+T_3\hat{\sigma}_z,\nonumber \\
&&\hat{T}_2=T_0+T_1\hat{\sigma}_x-T_3\hat{\sigma}_z,
\end{eqnarray}
with
\begin{eqnarray}\label{t}
&&T_0=t\,\sinh(x-i\xi)\,\cos\!{A},\nonumber \\
&&T_1=-i\,t\,\cosh(x-i\xi)\,\sin\!{A}\,\sin{\varphi},\nonumber \\
&&T_3=i\,t\,\sinh(x-i\xi)\,\sin\!{A}\,\cos{\varphi},
\end{eqnarray}
where we denoted
\begin{eqnarray}
t=\frac{\sinh{x}}{\sinh^2{(x-i\xi)}+\sin^2A\sin^2\varphi}.
\end{eqnarray}.

The reflection matrix $\hat{R}$ has the form:
\begin{eqnarray}\label{R}
&&\hat{R}_1=R_0+R_1\hat{\sigma}_x+R_2\hat{\sigma}_y,\nonumber \\
&&\hat{R}_2=R_0+R_1\hat{\sigma}_x-R_2\hat{\sigma}_y,
\end{eqnarray}
with
\begin{eqnarray}\label{r}
&&R_0=t\;\left[\coth
x\:\sin^2\!A\,\sin^2\!\varphi-i\frac{\sin\xi}{\sinh
x}\sinh(x-i\xi)\right],\nonumber \\
&&R_1=\frac{i}{2}\,t\;\sin\!{2A}\,\sin\varphi,\nonumber \\
&&R_2=\frac{i}{2}\,t\;\sin^2\!A\,\sin\!{2\varphi}.
\end{eqnarray}

Now we use Eqs.~(\ref{T}, \ref{t}) to obtain the transmission
probabilities, as the eigenvalues of the matrix $\hat{{\cal
T}}=\hat{T}^{\dagger}\hat{T}$:
\begin{eqnarray}\label{transmission_eigenvalues}
{\cal T}_{\pm}(\xi,x(\varphi))
=\frac{\sinh^2x}{\sinh^2x+\sin^2(\xi\pm\frac{\beta}{2})},
\end{eqnarray}
where the phase $\beta$ defined via
\begin{eqnarray}\label{beta}
\cos\beta=1-2 \sin^2\!\varphi\,\sin^2\!A
\end{eqnarray}
is due to the Rashba interaction.
Eq.~(\ref{transmission_eigenvalues}) demonstrates explicitly
spin-splitting of the transmission eigenvalues ${\cal T}_{\pm}$.
Note that $\beta=0$ and the splitting is absent for trajectories
with $\varphi=0$, which is the case of a purely 1-dimensional
(single-channel) contact~\cite{krive}. In the absence of normal
reflection, i.e. at $x \to \infty$, all transmission eigenvalues
are equal to unity and the spin-orbital effects disappear as
well~\cite{bezuglyi}.

\psfrag{T}{\kern-1pt\lower-2pt\hbox{{ $\cal{T}$}}}
\begin{figure}[tbp]
\includegraphics[angle=0,width=0.45\textwidth]{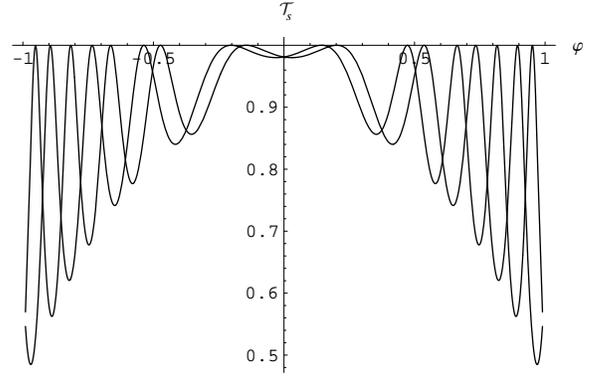}
\caption{\label{Transparency} The spin-splitted transmission
eigenvalues ${\cal T}_s$, $s=\pm 1$, as functions of the angle of
propagation $\varphi$ of the quasiparticles inside the 2DEG,
plotted for a realistic S-2DEG-S junction with parameters
$v_s=7\cdot 10^7 \text{cm/s}$, $v_n=5\cdot 10^7 \text{cm/s}$,
$m=m_e$, $m_n=0.035m_e$, $m_n \alpha/\hbar=5\cdot
10^4\text{cm}^{-1}$, $L=190 \text{nm}$. For these parameters and
for value of the superconducting gap $\Delta=1.4 \text{meV}$: (1)
the length of the contact $L$ is shorter than the coherence
length, $\xi_0=\hbar v_s/\Delta=330 \text{nm}$; (2) the Rashba
velocity is much smaller than the Fermi velocity in the 2DEG,
$\alpha/v_n\approx 0.03$; (3) the system is within the
semiclassical limit, $p_F L/\hbar=m_nv_n L/\hbar\approx 30$; (4)
the spin-orbital splitting $2\alpha p_F\approx 3.3\text{meV}$ is
larger than the superconducting gap $\Delta$; (5) the S/N
interfaces are almost transparent ($v_s/v_n\approx 1.4$), which
allows a large experimental value of the critical current.}
\end{figure}

Spin-orbital effect upon  ${\cal T}_{\pm}$ reduces, according to
Eq.~(\ref{transmission_eigenvalues}), to the shift of the main
semiclassical phase, $\xi \to \xi \pm \beta/2$, in agreement with
the result presented in Eq.~(1) of  Ref.~\onlinecite{Universal}.
An example of the dependence of transmission eigenvalues ${\cal
T}_{\pm}$ as functions of the incidence angle is shown in Fig.2.
An important point to mention is that this dependence is even with
respect to $\phi \to -\phi$ reflection, cf. Eq.(\ref{beta}). This
symmetry reflects a "trace" of the Kramers degeneracy which is
known to exist for the transmission eigenvalues defined within the
{\it real} basis of scattering states (note that the proof of
Kramers degeneracy of transmission eigenvalues is by far more
complicated~\cite{MelloPichard} than original Kramers theorem for
the degeneracy of energy levels). We characterize scattering
states by complex travelling waves $e^{ ip_y y}$, which leads to
breakdown of time invariance. Therefore  Kramers theorem is not
applicable to our model and spin-splitting demonstrated in
Eq.(\ref{transmission_eigenvalues}) may occur~\endnote{We are
grateful to C.W.J.Beenakker for the discussion of this point and
bringing to our attention Ref.~\onlinecite{MelloPichard}}.

It follows from Eq.~(\ref{transmission_eigenvalues}) that a
semiclassical average of any physical quantity which can be
expressed as a sum of terms containing individual ${\cal T}$
variables (i.e. does not contain cross-terms like ${\cal T}_+
{\cal T}_- $), does not depend upon the SO coupling. Indeed,
calculation of any average quantity in our model involves
integration over momentum component $p_y$ parallel to the
interfaces (or on propagation angle $\varphi$, defined as $p_y =
p_F\sin\varphi$). The integrand, as function of $\varphi$,
contains fast oscillations with characteristic scale $1/p_F L$,
and relatively slow dependence on $\cos\varphi$.  It is convenient
first to average over fast oscillations by going to probability
distribution of transmission eigenvalues defined as
\begin{eqnarray}\label{distribution_function}
{\cal P}_{\varphi}({\cal T}_{\pm})=\int \delta ({\cal T}-{\cal
T}_{\pm}(\xi,x(\varphi))\,d\xi.
\end{eqnarray}
Clearly, the presence of $\pm\beta$ phase-shift does not alter the
form of the probability distribution, which is of the same form as
considered, e.g. in Ref.~\onlinecite{MelBee} and independent on
the spin-orbit coupling:
\begin{equation}
{\cal P}_{\varphi}({\cal T})=\frac{\tanh{x}}{2 {\cal
T}\sqrt{1-{\cal T}}\sqrt{{\cal T}-\tanh^2{x}}}. \label{MB}
\end{equation}
Now we consider, as the simplest example, calculation of the
average conductivity of a junction in the normal state. It can be
written as
\begin{eqnarray}\label{average_conductivity}
G=G_Q\int_{-\pi/2}^{\pi/2}\cos\varphi
\frac{d\,\varphi}{\pi}\int{\cal
T}{\cal P}_{\varphi}({\cal T})\,d{\cal T}.
\end{eqnarray}
Universality of the distribution function ${\cal
P}_{\varphi}({\cal T})$ leads to independence of the average
conductance $G$, as well as of other quantities which can be
expressed via this distribution functions, upon the spin-orbit
phase $A$ (remember that we neglected weak effects of the order of
$\alpha/v_n \ll 1$ ). Note once again, that the above simple
considerations could not be applied to calculation of any quantity
which is {\it not additive} as a function of different
transmission channels, i.e. which contains products of different
transmission eigenvalues.

\begin{figure}
\includegraphics[angle=0,width=0.45\textwidth]{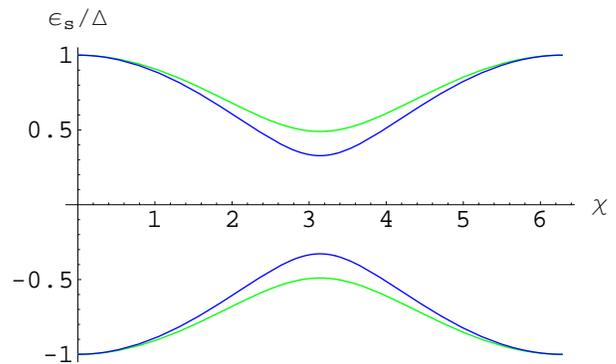}
\caption{\label{Andreev} The four spin-splitted Andreev levels
$\pm \epsilon_s$, $s=\pm 1$, as a function of the superconducting
phase difference $\chi$, plotted for a value of the angle of
propagation $\varphi=\pi/5$, and for a realistic S-2DEG-S
junctions with parameters $v_s=7\cdot 10^7 \text{cm/s}$,
$v_n=5\cdot 10^7 \text{cm/s}$, $\alpha\approx 0.2\cdot
10^7\text{cm/s}$, $m=m_e$, $m_n=0.035m_e$, $L=190 \text{nm}$.}
\end{figure}

\section{Josephson current}

Equation for the Andreev levels~(\ref{beenakker}) together with
Eq.~(\ref{transmission_eigenvalues}) for transmission eigenvalues
constitute the central result of the paper. Now one can calculate
Josephson current~\cite{Bee91}:
\begin{equation}
I(\chi) = \frac{e\Delta^2}{2\hbar}\sin\chi
\int\frac{L_ydp_y}{2\pi\hbar} \sum_{s=\pm 1}\frac{{\cal
T}_s(p_y)}{\epsilon_{s,+}(\chi)}\tanh\frac{\epsilon_{s,+}(\chi)}{2T}.
\label{Jos}
\end{equation}
Eq.~(\ref{Jos}) is applicable for the temperature-dependent
Josepshson current in the short-junction limit. In the
semiclassical limit $p_FL \to \infty$ the {\it average} Josephson
current can be calculated with the use of the distribution
function
 ${\cal P}_{\varphi}({\cal T})$
 given by Eq.~(\ref{MB}) as follows:
\begin{eqnarray}
\label{average_Josephson} I(\chi)=
\frac{e\Delta}{2\hbar}\int_{-\pi/2}^{+\pi/2}
\frac{d\,\varphi\cos{\varphi}}{\pi} \times \\ \nonumber \int {\cal
P}_{\varphi}({\cal T})\,d{\cal T} \frac{{\cal
T}\sin\chi}{\sqrt{1-{\cal T}\sin^2\frac{\chi}{2}}}
\tanh\frac{\Delta \sqrt{1-{\cal T}\sin^2\frac{\chi}{2}}}{2T}.
\end{eqnarray}
Eq.~(\ref{average_Josephson}) demonstrates independence of the
average Josephson current on the spin-orbital coupling. Such an
average current is a meaningful characteristic of a junction with
both lateral sizes much longer than the Fermi wavelength, $L, L_y
\gg \hbar/p_F$.

Oscillations of $I_c$ as a function of the electron
density, was discussed theoretically in
Ref.~\onlinecite{chrestin94} within the model very similar to the present one
(but without Rashba coupling).  It was argued that oscillations
should appear due to the presence of normal resonances in a
double-barrier structure, like the ones described in our
Eq.~(\ref{transmission_eigenvalues}) as a function of $\xi =
p_FL\cos\varphi$.  Strong spin-rotation effect expected
at $ L \geq L_0$, produces an additional phase $\beta \sim 1 $
which leads to splitting of the resonances as a function of
$p_FL$. As a result, at $ L \geq L_0$ oscillations discussed in
Ref.~\onlinecite{chrestin94} will have a twice shorter period
and  reduced amplitude.

\section{Spectrum equation and current
for the junction of an arbitrary length}

In this section we find  equations determining  the Andreev levels
for an arbitrary length of the contact, with the main purpose to
demonstrate, that the semiclassical average of the Josephson
current is independent upon the SO coupling for any $L/\xi_0$
ratio. Here we  use alternative method of calculation: instead of
expression of the Andreev levels via the transmission eigenvalues,
we employ  direct matching of wavefunctions obeying BdG equations
in the 2DEG and both superconductive regions. To simplify
calculations in this Section, we will consider here a model of
equal effective masses, $m_n=m$.

Eigenfunctions of the BdG equation for the S-2DEG-S junction can
be represented as 8-dimensional vectors, since they contain three
2-dimensional blocks: i) electron and hole components, ii) two
spin projections, and iii) two direction of momentum in the $x$
direction, $p_x =\pm p_F|\cos\varphi|$. Matching conditions for
wavefunctions on both S/N  interfaces consist of 16 scalar
equations which relate 8 wavefunction components in the 2DEG
region with 4 components in each of superconductive terminals (in
the case of the subgap Andreev levels which decay exponentially
into the bulk of the superconductors).  The next step is to reduce
this system of 16 equations down to 8 equations which couple
together 4+4=8 amplitudes of wavefunctions in superconductors. The
solvability condition for this system of 8 linear equations is
equivalent to the condition of vanishing of the corresponding
determinant,  $ g(\epsilon,\chi) =0$, which is equivalent to the
one defined in Eq.~(\ref{Andreevdet2}). After some tedious
calculations, the equation $g(\epsilon,\chi) =0$ can be
transformed into the following form:
\begin{eqnarray}
g(\epsilon,\chi) \equiv g_+(\epsilon,\chi)g_-(\epsilon,\chi) = 0 \, ,
\quad {\rm where} \nonumber \\
g_{\pm}(\epsilon,\chi)=\cos2\xi-Q\cos{\beta}\pm
\sqrt{1-Q^2}\sin{\beta} \, , \label{SpectrumEq}
\end{eqnarray}
where the parameter $\beta$ is defined in Eq.~(\ref{beta}),
\begin{eqnarray}
Q=\cos\Psi+\frac{4k^2K^2\Delta^2\left(\cos\Psi+\cos\chi\right)}
{\left(K^2-k^2\right)^2(\Delta^2-\epsilon^2)}, \quad {\rm and}
\nonumber \\
\Psi=2\arctan \frac{2kK
\epsilon}{(K^2+k^2)\sqrt{\Delta^2-\epsilon^2}}+ \mathcal{E},
\label{Q}
\end{eqnarray}
with $\mathcal{E}=2m \epsilon L/k$ being the energy dependent part
of the phase $\xi(\epsilon)$. Eqs.~(\ref{SpectrumEq}) demonstrate,
that in the presence of the Rashba interaction,  the Andreev
levels are generically spin-splitted for the contact of an
arbitrary length.

In the limit of vanishing spin-orbit interaction $\alpha=0$, as
well as for electron trajectories with $p_y=\varphi =0$, the
spectrum equation~(\ref{SpectrumEq}) reduces to the standard
equation $\cos2\xi=Q$ with a two-fold  degenerate (due to the
spin) solutions. In the special case of ideally transparent
boundaries $p_F=P_F$ general spectrum equation~(\ref{SpectrumEq})
also reduces to the standard equation $\cos2\xi=Q$, which now
simplifies to
\begin{equation}
\cos\left(-\mathcal{E}+2\arccos{\frac{\epsilon}{\Delta}}\right)=
\cos\chi.
\end{equation}

For a relatively short contact with $0 < \mathcal{E} \ll 1$ we expand
the Andreev spectrum equation~(\ref{SpectrumEq}) in powers of
small parameter $\kappa=\frac{m\Delta L}{k}$, and find the first
 correction to the result (\ref{beenakker}) obtained in Sec.II
for the $\kappa \to 0$ limit :
\begin{eqnarray}\label{SpectrumEqkappa}
\epsilon_{\pm}= \Delta\sqrt{1-{\cal
T}_{\pm}\sin^2\frac{\chi}{2}}\left(1-\kappa {\cal T}_{\pm}^{3/2}
\left|\sin\frac{\chi}{2}\right| \coth\,x\,\right),
\end{eqnarray}
where ${\cal T}_{\pm} $ are defined in
Eq.~(\ref{transmission_eigenvalues}).

In the general case of an arbitrary length of the contact, the
spectral equation (\ref{SpectrumEq}) is too complicated to be
solved explicitly for the energies of the Andreev levels.
Moreover, one should remember that for a junction of an arbitrary
$L/\xi$ ratio the continuous part of the spectrum (scattering
states) contributes to the Josephson current as well as the
localized levels we have considered. However, total Josephson
current (carried by both localized Andreev levels and continuous
part of the spectrum) can be found following the method presented
in Ref.~\onlinecite{nmsthesis}, in terms of the spectral function
$g(\epsilon,\chi)$ itself.

We will use Eqs.~(I.9), (A.48) and (A.49)  from
Ref.~\onlinecite{nmsthesis}, modified in our case due to the
presence of the spin-splitting, and the continuous scattering
channels characterized by the transverse momentum $p_y$. Therefore
the total current contains an integral over all $p_y$:
\begin{eqnarray}\label{nmsthesis1}
I_{\text{total}}(\chi)=L_y \frac{4e}{\hbar}T \sum_{s=\pm}
\int\frac{d p_y}{2\pi\hbar}\sum_{\omega_n>0}
\partial_{\chi}\ln g_s(i\omega,\chi),
\end{eqnarray}
here the summation goes over positive Matsubara frequencies:
$\omega=2\pi T(n+1/2)$, $n=0, 1,..$.

Within the semiclassical limit ($Lp_F\gg 1$), calculation of the
integral over $p_y$ in Eq.~(\ref{nmsthesis1}) can be simplified by
the same method which was used previously in the last part of
Sec.III. Namely, we first average over the period of the fast
oscillations of $\cos\xi \equiv \cos(kL)$ at a fixed angle
$\varphi$, and then do the integration over $\varphi$. Integration
$\int_0^{\pi} d\xi ...$ in Eq.(\ref{nmsthesis1}) leads to the
result which does not contain the spin-orbital phase $\beta$:
\begin{eqnarray}\label{Itotal}
I_{\text{total}}(\chi)= - L_y \cdot p_F\frac{4e}{\hbar^2}T
\sum_{\omega_n>0}\int_{-\frac{\pi}{2}}^{\frac{\pi}{2}}\,
\frac{d\varphi}{\pi}\cos{\varphi}\frac{\partial_{\chi}Q}
{\sqrt{|1-Q^2|}},
\end{eqnarray}
where $Q \equiv Q(\epsilon=i\omega_n,\chi)$ as defined in
Eq.~(\ref{Q}), and we took into account that $dp_y=p_F\cos\varphi
d\varphi$. Eq.~(\ref{Itotal}) demonstrates that the semiclassical
average of the Josephson current through the S/Rashba 2DEG/S
contact {\it does not depend} on the Rashba coupling constant, and
this result is valid for an arbitrary Fermi velocity mismatch and
arbitrary length of the contact. We note however that this result
is valid as long as electron-electron interaction in the 2DEG
region is neglected.

\section{Discussion}
The above results were obtained for a model of an infinitely long
junction in direction perpendicular to the current, when due to
the translational invariance the motion along $y$ axis was
completely determined by the wavevector $p_y$ of the corresponding
plane wave.  An obvious generalization of such a model system
would be a junction with periodic boundary conditions in  $y$
direction. In this case,  all our results would stay intact, up to
replacement of continuous $p_y$ by the discrete set of wavevectors
$p_n = 2\pi n/L_y$. Although being somewhat exotic for SNS
junctions, such a geometry does not seem to be impossible, taking
into account recent advances in fabrication of complicated InAs
structures, cf. e.g.~\cite{Arrays}. Usually, however,  S-2DEG-S
structure is of a finite length in $y$ direction $L_y$ with {\it
closed} boundary conditions,
 so that the channel eigenstates are
characterized by standing waves - mixtures of plane waves $e^{i
p_y L_y}$ and $e^{-i p_y L_y}$. In presence of the Rashba term,
the direction of the electron momentum is coupled to the direction
of its spin, thus determination of correct standing-wave
eigenstates is not trivial. The major effect of a finite $L_y \gg
L_0$ will be the existence of a discrete set of transmission
channels, $N_{ch} = 2L_y/\lambda_F$, where $\lambda_F$ is the
Fermi wavelength of the 2DEG.  However, some qualitative effect of
closed boundary conditions will occur: now one would deal with a
real basis of scattering states, thus Kramers theorem for the
transmission eigenvalues~\cite{MelloPichard} would be applicable.
It means that for closed boundary conditions and in the
short-junction limit $L/\xi_0 \to 0$, no spin-splitting of Andreev
levels may occur. In other terms, in a closed (in $y$ direction)
system Rashba coupling {\it modifies} transmission eigenvalues,
but does not {\it split} it. How can we reconcile it with a
natural idea that for very long $L_y$ the type of boundary
conditions should not be important? The point is that the total
Andreev spectrum of the system is doubly degenerate for the
periodic boundary conditions as well as for the closed ones. In
the first case, the degeneracy is due to the symmetry of ${\cal
T}_{\pm}(p_y)$ with respect to $p_y \to -p_y$ reflection, whereas
in the second case it is due to Kramers theorem. In order to get
global Andreev spectrum without degeneracy, the time-reversal
symmetry should be broken. In particular, it happens if nonzero
$L/\xi_0$ ratio is taken into account, as reported in
Ref.~\onlinecite{SN}. Another possibility might be related with an
{\it open} sample geometry like the one used in
Ref.~\onlinecite{5}, where additional current can be passed in the
direction transverse to the supercurrent.

Individual Andreev levels could possible be observed
experimentally by microwave spectroscopy or by measurement of
tunnelling conductance into the 2DEG region from an additional
point-like junction. One version of the former type of experiment
was proposed theoretically, for a single-channel junction, in
Ref.~\onlinecite{Lundin}. In this case, resonant frequency is very
high, of the order of $\Delta/\hbar$, since the only possible
transitions are between the positive and the negative Andreev
levels. This frequency is about 0.4 Thz  for Nb terminals
(considerably lower frequencies can be found in the case of a very
small reflection probability, $1-{\cal T} \ll 1$ and a phase
difference $\chi \approx \pi$). In many-channel junctions energy
spacing between neighboring Andreev levels is reduced as
$\delta\epsilon \sim \Delta/N_{ch}$, but usually (without
spin-orbit coupling)  it is not possible to observe
microwave-induced transitions between levels which belong to
different conduction channels. The reason is the momentum
conservation: different transmission channels are characterized by
different wavevectors $p_y/\hbar$ which are spaced by $\pi/L_y$,
whereas photon wavelength $\lambda_{ph} = hc/\delta\epsilon$ is
much longer than $L_y$, their ratio is of the order of
$(E_F^{2DEG}/\Delta)({c}/{v_n}) \sim 10^4$. It seems possible that
this selection rule will not be effective in the considered
situation with Rashba coupling, which modifies conduction channels
considerably  at $L_y \geq L_0$. The point is that now conduction
channels will be defined in the entangled space of orbital and
spin variables, thus there seems to be no reason for vanishing of
the inter-channel photon matrix element. However, this question
will certainly need further investigation.

\section{Conclusions}
We investigated the dependence of Josephson currents in clean
S-Rashba 2DEG-S proximity junctions upon the Rashba spin-orbit
interaction. We have generalized the Beenakker's formula for the
Andreev levels for the case of spin-orbital scattering and found
that for the  case of an infinitely wide junction (in direction
transverse to the current), the Andreev levels are spin-splitted.
This result is in agreement with  papers~[\onlinecite{bezuglyi, krive}], where
the effect of Rashba spin-orbital interaction upon supercurrent
was studied  either~\cite{bezuglyi} in the  case of absence of
normal backscattering at the interfaces ($p_F=P_F$), or in the
one-dimensional case~\cite{krive}.
We have shown that the semiclassical average of
the Josephson current is however insensitive to the Rashba
coupling as long as electron-electron interaction in 2DEG is
neglected.

Our results show therefore that the account of the Rashba
spin-orbital interaction for the usual model of SNS junction
without electron-electron interaction in the normal region is not
sufficient to explain experimentally observed strong suppression
of the $I_cR_N$ parameter with respect to its theoretical value.
We believe that to explain this suppression the electron-electron
interaction should be taken into account together with the
spin-orbital effects. We note that electron-electron interactions
in both density-density and spin-spin channels are not weak in the
2DEG structures.

A related  open problem is to find average spin polarization
$\langle S_y \rangle$ in the 2DEG region, which is expected to
exist at a nonzero supercurrent in the S-2DEG-S structure due to
symmetry considerations, cf. e.g. Ref.~\onlinecite{SN}. Note that
a supercurrent-induced average spin polarization will induce, in
the presence of electron-electron interaction, an effective Zeeman
field which may strongly modify the Andreev levels as well as the
Josephson current.

We are grateful to C. W. J. Beenakker, Ya. M. Blanter, I. V. Bobkova, A. M.
Finkelstein, F. Giazotto, P. M. Ostrovsky, N. M. Shchelkachev and
H. Takayanagi for many useful discussions. This research was
supported by the Dynasty Foundation and Landau-Juelich Scholarship
(O.D.), by RFBR grants 04-02-16348  and 04-02-08159,  by the
Programm "Quantum Macrosphysics" of Russian Academy of Sciences,
and by Russian Ministry of Education and Science via the contract
RI-112/001/417. Part of this research was accomplished during the
stay of O.Dimitrova at Laboratoire des Solides Irradies (Ecole
Polytechnique, Paris) within the scope of the ENS-Landau
collaboration programm.

\end{document}